\documentclass{aa} 
\usepackage{graphicx}
\usepackage{txfonts}
\usepackage{natbib}
\begin{document}
   \title{Rate and nature of false positives in the CoRoT exoplanet search \thanks{The CoRoT space mission, launched on December 27th 2006, has been developed and is operated by CNES, with the contribution of Austria, Belgium, Brazil , ESA (RSSD and Science Programme), Germany and Spain.}}

   \author{ J.M. Almenara \inst{1}
\and H.J. Deeg \inst{1}   
\and S. Aigrain  \inst{2}
\and R. Alonso \inst{3}
\and M. Auvergne \inst{4}
\and A. Baglin \inst{4}
\and M. Barbieri  \inst{3}
\and P. Barge \inst{3}
\and P. Bord\'e  \inst{5}
\and F. Bouchy \inst{6}
\and H. Bruntt \inst{7}
\and J. Cabrera \inst{8,9}
\and L. Carone \inst{10}
\and S. Carpano \inst{11}
\and C. Catala  \inst{4}
\and Sz. Csizmadia \inst{8}	
\and R. De la Reza \inst{12}
\and M. Deleuil \inst{3}    
\and R. Dvorak \inst{13}
\and A. Erikson \inst{8}	
\and M. Fridlund \inst{11}
\and D. Gandolfi \inst{14}
\and M. Gillon \inst{15,16}
\and P. Gondoin \inst{11}
\and E. Guenther \inst{14}
\and T. Guillot \inst{17}
\and A. Hatzes \inst{14}
\and G. H\'ebrard \inst{18}
\and L. Jorda \inst{3}
\and H. Lammer \inst{19}
\and A. L\'eger \inst{5}
\and A. Llebaria \inst{3}
\and B. Loeillet \inst{5,3}
\and P. Magain \inst{16}
\and M. Mayor \inst{15}
\and T. Mazeh  \inst{20}
\and C. Moutou \inst{3}
\and A. Ofir \inst{4}
\and M. Ollivier \inst{5}
\and M. P\"atzold \inst{10}
\and F. Pont \inst{2}
\and D. Queloz \inst{15}
\and H. Rauer  \inst{8,24}
\and C. R\'egulo \inst{1,21}
\and S. Renner \inst{8,22,23}
\and D. Rouan \inst{4}
\and B. Samuel \inst{5}
\and J. Schneider\inst{9}
\and A. Shporer \inst{20}
\and G. Wuchterl \inst{14}
\and S. Zucker \inst{20}
          }
\institute{
Instituto de Astrof\'isica de Canarias, C/ V\'ia L\'actea S/N, E-38200 La Laguna (Spain) \\\email{jmav@iac.es, hdeeg@iac.es}
\and School of Physics, University of Exeter, Stocker Road, Exeter EX4 4QL, United Kingdom
\and Laboratoire d'Astrophysique de Marseille, UMR 6110 CNRS, Technople de Marseille-Etoile, F-13388 Marseille cedex 13, France 
\and LESIA, UMR 8109 CNRS , Observatoire de Paris, UVSQ, Universit\'e Paris-Diderot, 5 place J. Janssen, 92195 Meudon, France
\and Institut d'Astrophysique Spatiale, UMR 8617 CNRS , bat 121, Universit\'e Paris-Sud, F-91405 Orsay, France       
\and Observatoire de Haute Provence, USR 2207 CNRS, OAMP, F-04870 St.Michel l'Observatoire, France
\and School of Physics A28, University of Sydney, Australia
\and Institute of Planetary Research, DLR, Rutherfordstr. 2, 12489 Berlin, Germany
\and LUTH, UMR 8102 CNRS, Observatoire de Paris-Meudon, 5 place J. Janssen, 92195 Meudon, France  
\and Rheinisches Institut f\"ur Umweltforschung, Universit\"at zu K\"oln, Abt. Planetenforschung, Aachener Str. 209, 50931 K\"oln, Germany
\and Research and Scientific Support Department, European Space Agency, ESTEC, 2200 Noordwijk, The Netherlands 
\and Observat\'orio Nacional, Rio de Janeiro, RJ, Brazil
\and Institute for Astronomy, University of Vienna, T\"urkenschanzstrasse 17, 1180 Vienna, Austria
\and Th\"uringer Landessternwarte Tautenburg, Sternwarte 5, 07778 Tautenburg, Germany
\and Observatoire de Gen\`eve, Universit\'e de Gen\`eve, 51 Ch. des Maillettes, 1290 Sauverny, Switzerland
\and University of Li\`ege, All\'ee du 6 ao\^ut 17, Sart Tilman, Li\`ege 1, Belgium
\and Observatoire de la C\^ote d'Azur, Laboratoire Cassiop\'ee, CNRS UMR 6202, BP 4229, 06304 Nice Cedex 4, France
\and Institut d'Astrophysique de Paris, UMR7095 CNRS, Universit\'e Pierre \& Marie Curie, 98bis Bd Arago, 75014 Paris, France
\and Space Research Institute, Austrian Academy of Sciences, Schmiedlstrasse 6, 8042 Graz, Austria
\and School of Physics and Astronomy, R. and B. Sackler Faculty of Exact Sciences, Tel Aviv University, Tel Aviv 69978, Israel
\and Dpto. de Astrof\'{i}sica, Universidad de La Laguna, La Laguna, 38206 Tenerife, Spain 
\and Laboratoire d'Astronomie de Lille, Universit\'{e} de Lille 1, 1 impasse de l'Observatoire, 59000 Lille, France
\and Institut de M\'{e}canique C\'{e}leste et de Calcul des Eph\'{e}m\'{e}rides, UMR 8028 du CNRS, 77 avenue Denfert-Rochereau, 75014 Paris, France
\and Center for Astronomy and Astrophysics, TU Berlin, Hardenbergstr. 36, D-10623 Berlin, Germany}

   \date{}

  
  \abstract
   {The CoRoT satellite {searches} for planets by applying the transit method, monitoring {up to} 12\,000 stars in the galactic plane for 150 days in each observation run. This search is contaminated by a large fraction of {false positives}, caused by different binary configurations that might be confused with a transiting planet.}
   {We {evaluate} the rates and nature of {false positives} in the CoRoT exoplanets search and compare our results with semiempirical predictions.}
   {We consider the detected binary and planet candidates in the first three extended CoRoT runs, and classify the results of the follow-up observations completed to verify their planetary nature. We group the follow-up results into undiluted binaries, diluted binaries, and planets and compare their abundances with predictions from the literature.}
   {{83\% of the initial detections are classified} as false positives using only the CoRoT light-curves, the remaining 17\% require follow-up observations. Finally, 12\% of the follow-up candidates are planets. The shape of the overall distribution of the {false positive} rate follows previous {predictions}, except for candidates with {transit depths} below about 0.4\%. For candidates with transit depths in the range from 0.1 - 0.4\%, CoRoT detections are nearly complete, and this difference from predictions is probably real and dominated by a lower than expected abundance of diluted eclipsing binaries.}
   {}


   \maketitle

%

\section{Introduction}

The CoRoT \citep{Baglin} Exoplanet Program is devoted to {a} planet search by the transit method. In each pointing it observes {up to} 12\,000 targets in different fields close to the Galactic center and anticenter. Detections by the transit method are contaminated to a large extent by `false positives', produced by sources other than transiting planets. Because of  {these false positives}, a sequence of tests - as originally outlined by \citet{Alonso} - is employed, beginning with {detailed} revisions of the detection light-curves, and continuing for surviving candidates with follow-up observations, to either reject them as planetary candidates or verify their planetary nature.
Results from these tests provide also in many cases insights into the nature of the {sources of the false positives}, which are generally produced by one of several kinds of configurations involving eclipsing binaries (EB). Hence, we obtain results about the population of these systems, which may be useful to the study of the distribution of binary systems, and for the calibration of the expected number and nature of {false positives} in upcoming and planned space missions for transit searches, {such as} Kepler \citep[for ground-based follow-up]{kepler,Gautier}, TESS \citep{Brown2008}, {and} PLATO \citep{plato}. 
 
\section{The sample of candidates}

CoRoT fields are located close to the Galactic plane. CoRoT obtains photometry for {up to} 12\,000 stars simultaneously, of  R-band magnitudes between {11} and 16. The photometry is obtained from onboard aperture photometry by means of large aperture masks, with a size and shape adapted to the large \emph{psf} of its exoplanet focal {plane}, where 50\% of the flux is contained in an elliptical area of about {35\arcsec x 23\arcsec}.

Previous ground-based observations characterized all the possible targets in each field \citep{exodat}; during the target selection, this allowed {preference to be given} to likely dwarf stars and hence to maximize the probability of finding a transit. For the brighter targets, CoRoT provides photometry in 3 different uncalibrated `colors' that can be used to reject planet candidates. 

\begin{figure}
   \centering
   \includegraphics[width=8.5cm]{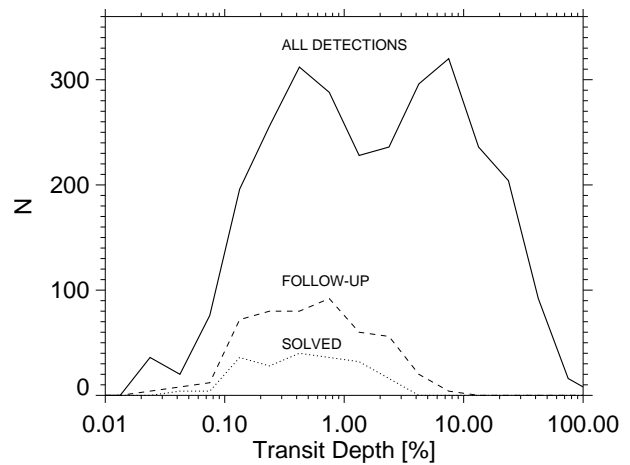}
   \caption{{Number of candidates found by the detection program, candidates in the observational follow-up program and candidates with solved follow-up observations. Candidate numbers are given in bins of a width of 0.25 log(transit depth); chosen in order to make this graph appear similar to Figs.~\ref{FigCoRoT} and \ref{FigBrown}.}}
              \label{FigCoRoT1}%
    \end{figure}

In our present study, we used data from the first three longer CoRoT observing runs, on which a significant part of the candidate follow-up was completed, i.e. the {`Initial Run' IRa01 ($\sim$55 days in the anticenter direction) and the first two `Long Runs',} lasting about 150 days each, which targeted the center and the anticenter (LRc01 and LRa01, respectively). We did not consider data from the Short Runs' since results from their candidate follow-up program remain largely incomplete. Several Detection Teams' studied CoRoT light-curves for planetary transits, also finding by coincidence a large {number} of {EBs}. A description of this detection effort and the properties of the entire sample will be given by \citet{Barge}. {Detection results are also described in more detail by \citet{Carpano} for the IRa01 observing run data and by \citet{Cabrera} for the LRc01 data, descriptions of the later runs being in preparation}. All detections from the light-curves (planets and {EBs}) are included in a common \emph{detection list}. Most of these detections are then rejected by means of an analysis of the light-curves, taking into account the depth, duration, shape and color signature, among others, and become classified as unspecified binary systems {\citep[for details of this procedure see][]{Carpano}}. The remaining candidates {are classified in terms of their ``planet-likeliness''}, and forwarded to the observational follow-up. For the three runs considered, this {amounts to} 122 candidates with different priorities. For each of {these} runs, Table \ref{detections} shows the numbers of observed targets, detections from the light-curves, candidates with follow-up programs, and candidates whose follow-up programs have finished. {We note that the ratio of candidates with follow-up programs to candidates with finished follow-up is approximately constant (see Fig.~\ref{FigCoRoT1}), without significant dependence on the transit depth, except for candidates with depths above 5\%, which were normally not selected for follow-up because they are obvious binaries.}

   \begin{table}
      \caption[]{Number of observed targets, detections, follow-up candidates and solved follow-up candidates in each field.}
         \label{detections}
     $$ 
         \begin{array}{p{0.25\linewidth}rrrrr}
            \hline \hline
            \noalign{\smallskip}
           & \#\; targets & \#\; det. & \%\; det. & \#\;cand. & \#\;solved   \\
            \noalign{\smallskip}
            \hline
            \noalign{\smallskip}
            IRa01          &  9872 & 230 & 2.33 & 39 & 19 \\
            LRa01          & 11408 & 299 & 2.62 & 45 & 6 \\
            Anticenter$^{\mathrm{*}}$     & 19878 & 499 & 2.51 & 80 & 24 \\
	    Center (LRc01) & 11408 & 226 & 1.98 & 42 & 25 \\
	    Total          & 31286 & 725 & 2.32 & 122 & 49 \\
            \noalign{\smallskip}
            \hline
         \end{array}
	 $$ 
\begin{list}{}{}
\item[$^{\mathrm{*}}$] Note that the two runs in the anticenter direction, IRa01 and LRa01 overlap partially, with 1402 targets, 30 detections, 4 candidates in follow-up and 1 solved follow-up candidate in common. 
\end{list}
   \end{table}

   \begin{table}
      \caption[]{Classification of the {solved} candidates from follow-up observations.}
         \label{follow-up}
     $$ 
         \begin{tabular}{p{0.3\linewidth}r|p{0.23\linewidth}r}
            \hline \hline
            Results from follow-up observations      &  \raisebox{-1.5ex}[0pt]{Number}  & \raisebox{-1.5ex}[0pt]{Classification}      &  \raisebox{-1.5ex}[0pt]{Number} \\
            \hline
            SB1   & 13 \\
            SB2   &  5 & \raisebox{1ex}[0pt]{Undiluted binary} & \raisebox{1ex}[0pt]{18}\\ 
            \hline
            Blend/Triple &  6 \\
            contaminating EB &  19 & \raisebox{1ex}[0pt]{Diluted binary} & \raisebox{1ex}[0pt]{25} \\ 
            \hline
            Planet &  6 & Planet   &  6 \\
            \hline
		Total & 49 & Total & 49\\
             \hline
         \end{tabular}
     $$ 
   \end{table}

\section{Classes of {false positives} from follow-up observations}
The main part of this study centers on an interpretation of the results from the observational follow-up, since {only} these results provide sufficient detail to {ascertain the nature of the false positives}.
 
\citet{Brown} completed the pioneering work on the types of {false positives} expected in transit searches. Discounting {false positive} sources based on observational or instrumental artifacts or produced by statistically possible coincidences of noise-features in low-amplitude detections, such sources are all related to eclipsing systems. The main types of {false positives} are then {EBs} that are observed directly ('undiluted binaries'), and {EBs} whose light is diluted by a nearby third star, which might be physically related to the system (triple star system) or be unrelated, with a third star being close to the line of sight to the binary system. Typically, this third star is brighter than the binary and corresponds to the observing target, whereas the binary is a faint background system. 
Only for low-amplitude candidates may we also have to consider \citep[e.g.][for the detection of {CoRoT-7b}]{Leger} a transiting system consisting of a star and a giant planet that is in the background and diluted by a brighter star, thereby mimicking a small-planet transit.
Among the undiluted binaries, planet-like eclipses may be caused by grazing EBs and the (central) eclipses of two stellar components with large ratios in area or surface brightness, typically due to a large mass ratio. {False positives caused by the transits of giants by main sequence stars are estimated to be negligible by Brown (see his Table 1).}

From the 122 candidates included in the observational follow-up, 49 of them were completely resolved at the time of {submission}, that is, 40\% of the total. The results of follow-up observations (see Table \ref{follow-up}) are a mixture of results from photometric follow-up \citep{Deeg}, which identifies contaminating' diluted EB's that are more than about 2\arcsec\ distant from the third star - which usually coincides with the observing target - , and from spectroscopic (radial velocity) results, which can only identify signals from sources that fall into the spectrograph's entry slit; that is, they have to be very close (less than 1-2\arcsec) to the target. Spectroscopy may thereby identify spectroscopic blends, which correspond to diluted EBs (these may be bound triple systems or be unbound), or undiluted EBs in the form of single or double-lined spectroscopic binaries (SB1, SB2), or lastly, extrasolar planets.

Since the physical nature of the diluted systems can often not be established, we have used in this analysis only three classifications: undiluted binaries, diluted binaries, and planets, as shown in Table \ref{follow-up}. We note that a candidate-by-candidate description of results from the follow-up is given for IRa01 by \citet{Moutou} {and for LRc01 by \citet{Cabrera}}; for the other runs,  similarly detailed descriptions will be published once follow-up has been completed.

   \begin{figure}
   \centering
   \includegraphics[width=8.5cm]{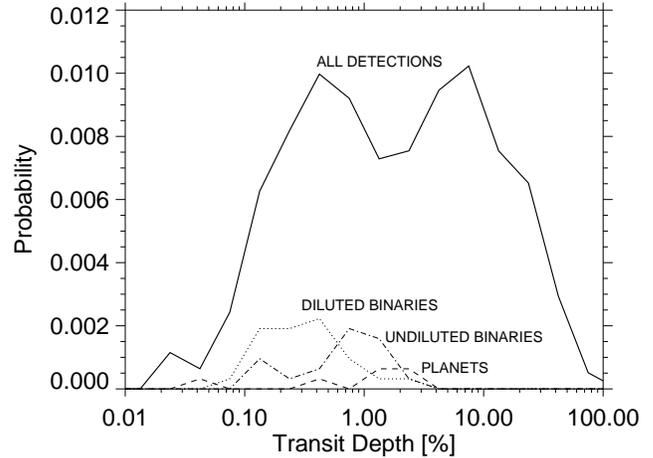}
   \caption{Probability of occurrence of the classes of {false positives} detected by CoRoT, per unit log of transit depth, {based on} the amplitude found in the original CoRoT light-curve. The classifications based on observational follow-up have been multiplied by 2.5; see text. The upmost line indicates the distribution of all candidates found by the detection program, {identical to Fig.~\ref{FigCoRoT1}}.}
              \label{FigCoRoT}
    \end{figure}

\section{Discussion}
Figure~\ref{FigCoRoT} shows the probability of the occurrence of the various classifications among CoRoT targets, per unit $\log$ of transit depth. It also shows the distribution of all the candidates found by the detection effort (independent of whether they were selected for observational follow-up or not).

   \begin{figure}
   \centering
   \includegraphics[width=8.5cm]{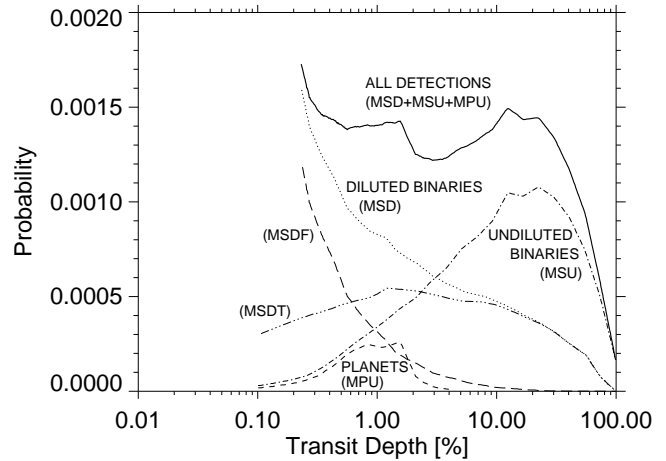}
   \caption{Adaptation of Figure 3 from \citet{Brown}, of probability estimates for a ground based survey in a field in Cygnus. In brackets the notation from Brown. Axes are similar to Fig.~\ref{FigCoRoT}.} 
              \label{FigBrown}
    \end{figure}

In comparison, we show in Fig.~\ref{FigBrown} {semi-empirical} estimates of these probabilities calculated by \citet{Brown} for a ground-based small telescope transit search on a somewhat brighter sample of 9 $\leq$ R $\leq$ 12 in a region in Cygnus. We select these estimates instead of newer ones from \citet{Brown2008} made for the TESS satellite, because \citet{Brown} considers a region in the Galactic plane similar to CoRoT, whereas \citet{Brown2008} integrate over the entire sky. In the original work by Brown, slightly different classes of {false positives} were used: MPU (main-sequence star with a giant planet); MSU (undiluted binaries); and the two types of diluted binaries, MSDF (an eclipsing binary + a third non-related star) and MSDT (triple systems). In Fig.~\ref{FigBrown}, we have summed these {last two} types into a group called 'diluted binaries'. Among the undiluted binaries, we note that Brown only mentions grazing binaries as a principal source of {false positives}; however, as can be seen in Table \ref{follow-up}, eclipses among stellar components with large {area} or surface-brightness ratio (SB1 in Table \ref{follow-up}) are the cause of a significant fraction of {false positives}. 

As for Brown, the summed distribution of CoRoT candidates exhibits a peak around 10\% eclipse depth, probably also caused by the population peak of undiluted binaries. The high-amplitude region of depths over 5\% is not covered by candidates in the follow-up program, because the eclipse depth as such identifies these cases as {false positives} in planet finding, excluding them automatically from follow-up observations. 

Classifications of the type of {false positive} (indicated by the lower lines\footnote{Probabilities for the results from follow-up observations have been multiplied by a factor of 2.5, which corresponds to the ratio of the 122 candidates in the program to the 49 {solved} ones; here we assume that the unresolved candidates follow the same distribution as the resolved ones.} in Fig.~\ref{FigCoRoT}) could be derived only for candidates resolved from follow-up observations. In these classifications, we note that the distributions of 'diluted binaries' and 'undiluted binaries' from CoRoT follow the same trends as in Brown down to transit depths of about 0.4\%, below which the overall candidate population begins to decline. This decline is probably a true property of the CoRoT sample, and not caused by failures in detecting existing candidates, at least in the zone of depths between 0.1\% and 0.4\%. Here, candidate detections should be nearly complete; with incomplete detection returns only dominating below amplitudes of 0.1\% \citep{Barge}. We instead expect that there are fewer faint diluted binary systems (which cause these low-amplitude candidates) {than} assumed in the simulations by Brown. Results from the photometric follow-up of CoRoT candidates \citep{Deeg} show that most {of the} diluted binaries that have been identified from low-amplitude (0.1 - 0.5\%) candidates fall into the magnitude range of 16-19. {Differences between the extinction of that population and that of the corresponding one in Brown, which would be of the order of 12-15 mag, and in the real-versus-simulated fraction of binary stars, are most likely the causes for these different trends. Brown assumes a constant fraction of 0.49 for main-sequence stars. We note that the population of background stars in CoRoT is $\sim$3 magnitudes fainter than those of Brown. In consequence, this population will typically be of later spectral types, which presents lower fractions of binary systems \citep{Lada}; hence the impact of false alarms from faint EBs is less for deeper samples. Thus, the sharp rise toward shallower transits predicted by Brown for diluted binaries is an overestimate, as observed by CoRoT.}

The relatively small size of the sample of candidates in the follow-up program illustrates the efficiency of {false positive} rejections based solely on the light-curves from CoRoT. These procedures use combinations of transit depth, duration, shape, and the presence of color signatures for the selection of candidates worthy of further follow-up; however, a classification into {false positive} types based on these light-curve rejections has not been possible since many of the criteria employed do not provide this information. However, we note that the fraction of  `all detections' to candidates in the follow-up program' is {approximately} constant below transit depths of 3\% (see Fig.~\ref{FigCoRoT1}).

Summing up the 'all detections' probabilities from Brown that are deeper than 0.23\% {(corresponding to the lower limit to which this can be done from Brown's graph and} the CoRoT sample can be considered almost complete), we derive an overall {predicted} detection rate of 0.34\%, whereas from CoRoT data we obtain a 1.88\% {actual} detection rate, or 5.5 times higher than that of Brown. The estimations of Brown were made for a planet transit search centered on the Galactic plane in Cygnus (matching the Kepler field) for a sample of stars brighter than 12, and Brown considers these estimates to be valid to within a factor of 2 for the ground-based surveys that were the subject of his study. Factors affecting that difference apart from those already mentioned are: the different angular distance from a target within which EBs become diluted and cause {a false positive}; here Brown uses a radius of 20\arcsec\, whereas for CoRoT, diluted EBs found by the photometric follow-up infer a radius of {about 17\arcsec \citep{Deeg}}.{The area is then 1.4 times larger in Brown than for CoRoT}. Furthermore, the Brown sample has a lower dwarf-to-giant ratio, {being about one third}. This is because of the brighter stellar sample that he uses, and from which he considers \emph{all} stars, whereas CoRoT targets are being assigned after a preselection that rejects giant stars, resulting in a fraction of dwarfs of about 67\%; this difference affects the undiluted binary and planet distributions, {that is, they should be a factor of 2 higher in CoRoT than in Brown. The stellar density of the contaminants, dominated in CoRoT by stars 3 magnitudes fainter than the faintest target \citep{Deeg}, is an estimated factor of 9 times larger in CoRoT than for the sample analyzed by Brown. This factor is estimated from the counts of faint stars in the CoRoT fields \citep[Fig. 7 in][]{exodat}, comparing them at the dominant magnitude for both contaminants in CoRoT and the sample analyzed by Brown. A higher contaminant density can be expected to linearly increase the fraction of target stars that are contaminated by faint background binaries.}  

Finally, the planet-distribution of \citet{Brown} is one of hot giant planets; CoRoT's results contain to date 5 hot giants which follow approximately the distribution of Brown, and one terrestrial planet ({CoRoT-7b}), visible as a small peak at a transit depth of 0.04\% in Fig.~\ref{FigCoRoT}. {The hot giant yield is close (less than a factor of 2) to the expected one based on ground-based transit surveys. The current lack of small planet detections is likely to correspond to underabundances of these objects, as  noted previously \citep{Mazeh,Southworth} and probably corresponds to a lower limit to the existence of gaseous planets because of evaporation.  \citet{Davis} present evidence of this limit that can be described as a line in the $M_p^2/R_p^3$ versus $a^{-2}$ plane (with $a$ being the orbital distance). Short-periodic Neptune-like planets are close or below this limit, and consequently transform into smaller and denser planets, whose transit detection still eludes us in most cases, CoRoT-7b being the first possible member of the remaining population of nuclei of gas-planets that have undergone significant evaporation.}

\section{Conclusions}
Candidate detections have been obtained {for} about 2.3\% of all CoRoT targets. Of these, {a very high fraction} (83\% ) has been rejected as planet candidates {solely on the basis of an analysis of the discovery light-curve, and only 17\% needed follow-up observations to clarify their planetary or non-planetary nature. Of those targeted by follow-up observations, about 12\% of candidates were {verified as} planets. {For the first three runs described here, the selection of the candidates for the follow-up observations was based on very wide criteria aimed to avoid the rejection of valid candidates, and only clearly unsuitable candidates were rejected from the light-curve analysis.  Among these 83\% of rejected cases, were mostly binaries with clear primary and secondary eclipses, or with primary eclipses of depths greater than 5\%. Furthermore, cases with clear differences among eclipses in the three CoRoT passbands were rejected. The results from the follow-up of the first runs, as described here, have also shown that none of the candidates with low planet likeliness' turned out to be a planet.  In later CoRoT-runs, this has led to a more stringent rejection of candidates from the light-curve analysis, thereby reducing the observational load from numerous candidates with low planet-likeliness.}

Since we cannot determine the true nature of \emph{all} the CoRoT detections, -- only the non-planet status is known for candidates rejected on the basis of light-curve analysis alone -- we cannot directly compare with the predictions from \citet{Brown}. However, we can evaluate the overall distribution of candidates, as well as the trends for those that have been classified from follow-up observations.

Regarding the overall distribution, the CoRoT candidates with amplitudes of 0.5\% or larger closely follow that of Brown, except for being lower by a scale factor of about 5.5. CoRoT candidate abundances in the range of 0.1-0.4\% are significantly lower, probably due to an observed trend indicating fewer diluted binaries than {predicted}, where as for even lower amplitudes, low detection efficiencies begin to dominate.   

The lower abundance of diluted binaries between 0.1 and 0.4\% transit depth, which possibly continues towards lower-amplitude candidates, may facilitate the transit-searches for small extrasolar planets in both CoRoT data and in upcoming or planned search projects, such as Kepler or PLATO, reducing the load of required follow-up observations.

When more candidates have become observed in follow-up, it will be very interesting to divide the CoRoT sample into candidates in the center and anticenter fields, which have different stellar densities; this should provide more detailed insight into the  abundances of diluted and undiluted binaries. Also, the distribution of planet detections, currently affected by small number statistics, should become better established.

\begin{acknowledgements}
J.M.A. and H.J.D. acknowledge support by grants ESP2004-03855-C03-03 and ESP2007-65480-C02-02 of the Spanish Ministerio de Educaci\'on y Ciencia. 
T.M. acknowledges the supported by the Israeli Science Foundation (grant no. 655/07). 
\end{acknowledgements}

\end{document}